\documentclass[aps,preprint,amsmath,amssymb,tightenlines]{revtex4}
\usepackage{graphicx,color,slashed}
\usepackage{subfigure}
\def\be{\begin{eqnarray}}
\def\ed{\end{eqnarray}}

\begin{document}

{\begin{flushright}{KIAS-P15057}
\end{flushright}}

\title{  SU(2)$_L \times$SU(2)$_R$ minimal dark matter with 2 TeV $W'$  }

\author{ \bf P. Ko$^{a}$\footnote{Email: pko@kias.re.kr} and Takaaki Nomura$^{a}$\footnote{Email:
nomura@kias.re.kr} }

\affiliation{ $^{a}$ School of Physics, Korea Institute for Advanced Study, Seoul 130-722, Republic of Korea}

\date{\today}

\begin{abstract}
We construct the minimal dark matter models in the left-right symmetric extensions of the
standard model (SM), where the gauge symmetry SU(3)$_C \times$SU(2)$_L \times$
SU(2)$_R \times$U(1)$_{B-L}$ is broken into its subgroup SU(3)$_C\times$ 
U(1)$_{\rm em}$ by nonzero VEVs of a SU(2)$_R$ doublet $H_R$ and a SU(2)$_L \times$ SU(2)$_R$ bidoublet $H$.
A possible candidate of dark matter is explored in the framework of minimal dark matter considering SU(2)$_{L, R}$ multiplet scalar bosons and fermions.
Then we focus on SU(2)$_R$ quintuplet fermions with $B-L$ charges 0, 2 and 4 as the minimal dark matter candidates and investigate phenomenology of them.  
We show that the dark matter in the model can provide observed relic density with 2 TeV $W'$ boson which is motivated by the ATLAS diboson excess and CMS $eejj$ excess.
The possible mass of dark matter is predicted for each $B-L$ charge.
We then estimate the scattering cross section of dark matter with nucleon and production cross section of charged components in the quintuplets at the LHC.

\end{abstract}

\maketitle

\section{Introduction}
 
The Standard Model (SM) based on SU(3)$\times$SU(2)$_L \times$U(1)$_{Y}$ has been 
very successful in describing particle physics phenomena from low energy to $\sim$ (a few) 
TeV.  Still it suffers from a number of phenomenological drawbacks such as neutrino masses 
and mixings, nonbaryonic dark matter (DM) and matter-antimatter asymmetry as well as 
cosmological inflation, not even mentioning theoretical puzzles such as fine tuning of Higgs 
mass, the origin of flavor and generation, and strong CP problem, etc..  There are a number 
of different extensions of the SM, based on which problem it aims to solve.  

One interesting gauge extension of the SM would be left-right (LR) symmetric models, where 
the right-handed fermions form SU(2)$_R$ doublets, similarly to the left-handed (LH) 
fermions forming SU(2)$_L$ doublets~\cite{Mohapatra:1974gc,Pati:1974yy,Senjanovic:1975rk,Senjanovic:1978ev,Mohapatra:1979ia,Mohapatra:1980yp}. In this case one can enjoy a possibility that the left-right symmetry is completely 
restored in high energy scale if we assume $g_L = g_R$ and similar assumptions for Yukawa couplings. 
Also strong CP problem~\cite{Mohapatra:1997su, Gu:2010zv}, neutrino masses and mixings, and lepton number violation~\cite{Keung:1983uu, Maiezza:2010ic,Tello:2010am, Guadagnoli:2011id, Nemevsek:2012iq, Chen:2013fna, Dev:2013oxa} can be addressed as well from different 
viewpoints.  

The canonical version of the LR model is to assume the exact left-right symmetry with the following 
Higgs sector, SU(2)$_L$ triplet  $\Delta_L$,  SU(2)$_R$ triplet $\Delta_R$ and SU(2)$_L \times$ 
SU(2)$_R$ bidoublet $\phi$.  Then, the LR symmetry is assumed to be spontaneously broken into 
$U(1)_{\rm em}$ by nonzero VEVs of  $\Delta_R$ and $\phi$.   This setup has been 
studied extensively, including flavor physics~\cite{Keung:1983uu, Maiezza:2010ic,Tello:2010am, Guadagnoli:2011id, Das:2012ii, Deppisch:2014zta, Rodejohann:2015hka, Awasthi:2015ota, Nemevsek:2012iq, Chen:2013fna, Dev:2013oxa}, minimal DM~\cite{Heeck:2015qra} and right-handed neutrino DM~\cite{Bezrukov:2009th, Nemevsek:2012cd}.  

However the assumption of exact LR symmetry may be a too strong and constraining 
assumption.  In principle, the SU(2)$_L$ and SU(2)$_R$ gauge couplings could be different even 
at high energy scales. Moreover one can achieve the necessary gauge symmetry breaking 
with nonzero VEVs of SU(2)$_R$ doublet $H_R$ instead of triplet $\Delta_R$ and a 
bidoublet $\phi$, which is much simpler than the canonical LR models.  
We also note that inverse seesaw mechanism can be applied introducing gauge singlet fermions~\cite{Wyler:1982dd, Mohapatra:1986bd, GonzalezGarcia:1988rw}.

In fact, this simpler setup including SU(2)$_L$ doublet Higgs has been studied recently~\cite{Deppisch:2015cua}, mainly motivated by the 
ATLAS  2 TeV diboson excess~\cite{Aad:2015owa} where a moderate excess is also found by CMS~\cite{Khachatryan:2014hpa, Khachatryan:2014gha}.
The 2 TeV $W'$ boson from SU(2)$_R$ is a possible explanation of the excess, which is also 
discussed in the canonical LR models~\cite{Awasthi:2015ota, Heeck:2015qra, Hisano:2015gna, Cheung:2015nha, Dobrescu:2015qna, Aguilar-Saavedra:2015rna, Gao:2015irw, Brehmer:2015cia, Cao:2015lia, Dev:2015pga}.  
Although one has to await more data accumulation at LHC Run 2,  it is interesting to ask oneself 
what kind of new physics may explain this tantalizing  ATLAS 2 TeV diboson excess, 
if it is a real signature of physics beyond the SM.
Furthermore, there is also some excess in the $eejj$ channel of the CMS search~\cite{Khachatryan:2014dka}, which could be also discussed 
as a potential signal of $W'$ decaying into electron and right-handed neutrinos~\cite{Dev:2015pga,Coloma:2015una,Deppisch:2015cua, Heikinheimo:2014tba,Gluza:2015goa}.

In this letter we construct a minimal dark matter model in the SU(2)$_L \times$ SU(2)$_R 
\times$ U(1)$_{B-L}$ extension where the gauge symmetry is broken down to U(1)$_{\rm em}$ 
by nonzero VEV's of a bidoublet $H$ and a SU(2)$_R$ doublet $H_R$ \footnote{Recently Heeck et al. also studied the minimal DM in the canonical LR models~\cite{Heeck:2015qra}.  
The allowed DM mass range in their setup is heavier than what we find out in this paper, and 
DM phenomenology is qualitatively different depending on the details of the Higgs sector.}.
We shall explore the possible candidates of DM introducing SU(2)$_{L, R}$ multiplet scalar bosons 
and fermions based on the idea of minimal DM~\cite{Cirelli:2005uq}. 
Then the SU(2)$_R$ quintuplet fermions with $B-L=$ 0, 2 and 4 are investigated as candidates of minimal DM of SU(2)$_R$ multiplet.
We show that properties of the minimal DM such as the mass splitting within the multiplet are 
different from that in the canonical LR models~\cite{Heeck:2015qra}.
The allowed DM mass range is then different in two cases, and so is the resulting phenomenology. 
Particularly we find that our minimal DM can be accommodated with 2 TeV $W'$ where the relic density tends to be smaller than observed value in the canonical LR model for the parameter region in which a neutral component of SU(2)$_R$ multiplet is the lightest one because of different mass splitting pattern. 
The mass values of DM are predicted for $m_{W'}=2$ TeV. We then estimate DM-nucleon scattering cross section for DM direct detection and production cross section of charged components in the DM  multiplet at the LHC Run 2.

\section{The model}

Let us consider a left-right extension of the SM applying the gauge symmetry SU(3)$\times$SU(2)$_L \times$SU(2)$_R \times$U(1)$_{B-L}$ where $B$ and $L$ are baryon and lepton numbers respectively~\cite{Mohapatra:1974gc,Pati:1974yy,Senjanovic:1975rk,Senjanovic:1978ev,Mohapatra:1979ia,Mohapatra:1980yp}.
The scalar contents are taken to be 
\begin{align}
H: (2,\bar{2},0), \quad H_R: (1,2,1)
\end{align}
where the quantum numbers in the parenthesis are under the SU(2)$_L \times$SU(2)$_R \times$U(1)$_{B-L}$ omitting SU(3) for simplicity.
Here we have loosened the exact left-right symmetry and applied more simplified choice of the scalar contents to break the gauge symmetry.
The fermion contents are also given by
\begin{align}
& q_L : (2,1,1/3), \quad q_R : (1,2,1/3), \quad \ell_L : (2,1,-1), \quad \ell_R : (1,2,-1), \quad S:(1,1,0),
\end{align}
where Majorana fermion $S$ is introduced to lead inverse seesaw mechanism~\cite{Wyler:1982dd, Mohapatra:1986bd, GonzalezGarcia:1988rw}.

\begin{figure}[t] 
\begin{center}
\includegraphics[width=60mm]{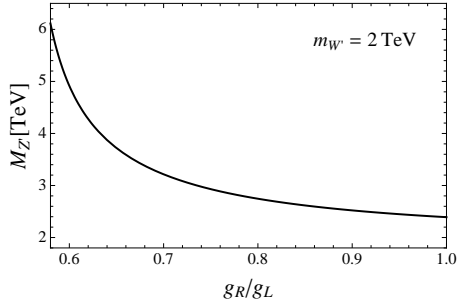}
\caption{ Mass of heavy neutral gauge boson $Z'$ as a function of $g_R/g_L$ with $m_{W'} = 2$ TeV.}
\label{Fig:mZ'}
\end{center}
\end{figure}

The gauge symmetry is broken by non-zero vacuum expectation value of $H$ and $H_R$. We assume the VEVs of the scalar fields are developed such that 
\begin{equation}
\langle H \rangle = \frac{1}{\sqrt{2}} \begin{pmatrix} k_1 & 0 \\ 0 & k_2 \end{pmatrix}, \, \langle H_R \rangle = \begin{pmatrix} 0 \\ v_R/\sqrt{2} \end{pmatrix}.
\end{equation}
The gauge symmetry is then broken as SU(2)$_L \times$SU(2)$_R \times$U(1)$_{B-L} \to$U(1)$_{em}$ where the electric charge $Q$ is given by 
\begin{equation}
Q = T^3_L + T^3_R + \frac{1}{2} Q_{B-L}
\end{equation}
where $T^3_{L(R)}$ and $Q_{B-L}$ denote the diagonal generator of SU(2)$_{L(R)}$ and the $B-L$ value of a field respectively. 
After spontaneous symmetry breaking, we obtain the mass terms of the gauge bosons such that
\begin{align}
L_M =& (W_L^{+ \mu}, W_R^{+ \mu}) \tilde{M}_W^2 \begin{pmatrix} W_{L \mu}^- \\ W_{R \mu}^- \end{pmatrix}+ \frac{1}{2} (W^{3 \mu}_{ L}, W^{3 \mu}_R, X^\mu) \tilde{M}_0^2 \begin{pmatrix} W^3_{L \mu} \\ W^3_{R \mu} \\ X_\mu \end{pmatrix},
\end{align}
where $W^{\pm \mu}_{L,R} = (W^{1 \mu}_{L,R} \mp i W^{2 \mu}_{L,R})/\sqrt{2}$. The mass matrices are given by
\begin{align}
\tilde{M}_W^2 = \frac{1}{4} \begin{pmatrix} g_L^2 K^2 & - 2 g_L g_R k_1 k_2 \\ -2 g_L g_R k_1 k_2 & g_R^2 (K^2 + v_R^2) \end{pmatrix}, \,
\tilde{M}_0^2  =  \begin{pmatrix} \frac{g_L^2}{4} K^2 & - \frac{g_L g_R}{4} K^2 & 0 \\ -\frac{g_L g_R}{4} K^2 & \frac{g_R^2}{4} (K^2 +  v_R^2) & \frac{g_R g_{B-L}}{4} v_R^2, 
 \\ 0 &  \frac{g_R g_{B-L}}{4} v_R^2 & \frac{g_{B-L}^2}{4} v_R^2 \end{pmatrix}
\end{align}
where $K^2 = k_1^2 + k_2^2$ and $g_{L,R}$ and $g_{B-L}$ are gauge couplings of 
SU(2)$_{L, R}$ and U(1)$_{B-L}$.
The mass matrices are diagonalized by the orthogonal transformation~\cite{Duka:1999uc}
\begin{align}
\label{eq:mixing1}
\begin{pmatrix} W_L^\pm \\ W_R^\pm \end{pmatrix} &= \begin{pmatrix} \cos \xi & \sin \xi \\ - \sin \xi & \cos \xi \end{pmatrix} \begin{pmatrix} W^\pm \\ W'^\pm \end{pmatrix}, \\
\begin{pmatrix} W_{3L} \\ W_{3 R} \\ X \end{pmatrix} &= \begin{pmatrix} c_W c_X & c_W s_X & s_W \\ - s_W s_M c_X -c_M s_X & -s_W s_M s_X + c_M c_X & c_W s_M \\ -s_W c_M c_X + s_M s_X & -s_W c_M s_X -s_M c_X & c_W c_M  \end{pmatrix} \begin{pmatrix} Z \\ Z' \\ A \end{pmatrix},
\end{align}
where $s_W(c_W) = \sin \theta_{W}(\cos \theta_W)$ with the Weinberg angle $\theta_W$, $s_M \equiv \sin \theta_M = g_{B-L}/\sqrt{g_R^2 + g_{B-L}^2}$, $c_M \equiv \cos \theta_M = g_R/\sqrt{g_R^2 + g_{B-L}^2}$, 
and $s_X(c_X) = \sin \theta_X (\cos \theta_X)$ is associated with mixing of massive neutral gauge bosons.
The mixing angles $\xi$ and $\theta_X$ are assumed to be very small and will be ignored 
in the following analysis of dark matter phenomenology.
The gauge couplings satisfy the relations $g_R = e/(s_M c_W)$ and $g_{B-L} = e/(c_W c_M)$.
The $c_M$ and $s_M$ can be rewritten as 
\begin{align}
s_M = \tan \theta_W \left( \frac{g_R}{g_L} \right)^{-1}, \quad
c_M = \left( \frac{g_R}{g_L} \right)^{-1} \sqrt{\left( \frac{g_R}{g_L} \right)^{2} - \tan^2 \theta_W } \, , 
\end{align}
where the gauge coupling should satisfy $g_R/g_L > \tan \theta_W$ for consistency.
Assuming $K \ll v_R$, the mass eigenvalues of new gauge bosons are approximately given by
\begin{align}
m_{W'}^2 & \simeq \frac{1}{4} g_R^2 v_R^2 \left( 1 + \frac{K^2}{v_R^2} \right), \\
m_{Z'}^2  & \simeq \frac{1}{4} (g_R^2+ g_{B-L}^2 ) v_R^2 \left( 1 + \frac{K^2 c_M^2}{v_R^2} \right)  .
\end{align}
Thus the mass relation of heavier gauge bosons are approximately
\begin{equation}
\frac{m_{Z'}}{m_{W'}} \simeq \frac{g_R/g_L}{\sqrt{(g_R/g_L)^2 - \tan^2 \theta_W}} \ .
\end{equation}
Fig.~\ref{Fig:mZ'} shows the $m_{Z'}$ as a function of $g_R/g_L$ with $m_{W'}=2$ TeV where  $m_{Z'}/m_{W'} \sim 1.2$ for $g_L = g_R$.
We note that extra factor of $\sqrt{2}$ appear in the numerator of RHS when an SU(2)$_R$ triplet develops VEV instead of doublet. 

The left-right related $W'$ boson with $m_{W'} \sim 2$ TeV would explain the diboson excess observed in the ATLAS experiment~\cite{Aad:2015owa} where the $W'$ is produced as $\bar q' q \rightarrow W'$ and decays into $WZ( \rightarrow jj)$ at the LHC~\cite{Awasthi:2015ota, Heeck:2015qra,Hisano:2015gna, Cheung:2015nha, Dobrescu:2015qna, Aguilar-Saavedra:2015rna, Gao:2015irw, Brehmer:2015cia, Cao:2015lia, Dev:2015pga, Deppisch:2015cua, Goncalves:2015yua}. Furthermore 2 TeV $W'$ also can be an explanation of $eejj$ excess in the CMS search~\cite{Khachatryan:2014dka} when $W'$ decays electron and right-handed neutrino~\cite{Dev:2015pga,Coloma:2015una,Deppisch:2015cua, Heikinheimo:2014tba, Gluza:2015goa}. It is also indicated that coupling relation of $g_R < g_L$ is suitable to explain these excesses~\cite{Heikinheimo:2014tba, Dev:2015pga,Deppisch:2015cua}.   Motivated by these possibility, we fix the mass of $W'$ as 2 TeV and consider two different cases, 
$g_R/g_L =1.0$ and $g_R/g_L = 0.6$  in following analyses.

To accommodate minimal dark matter (DM) within our model, we consider a new fermion or scalar 
SU(2)$_{L, R}$ multiplet.
If a multiplet does not have interaction leading its decay up to dimension-6 operator level the lightest component can be stable in cosmological time-scale by the same idea as Minimal Dark Matter (MDM) model~\cite{Cirelli:2005uq}. 
In Tables.~\ref{tab:DM}, we summarize possible interactions for decay of SU(2)$_{L,R}$ multiplet scalar(fermion) $\Phi_{L,R}(\Psi_{L,R})$ up to quintuplet. 
We find that scalar SU(2)$_L$ quintuplet with non-zero $Q_{B-L}$ does not have an operator less 
than dimension 6 while scalar SU(2)$_R$ quintuplet have dimension 5 operator.
On the other hand,  fermion SU(2)$_L$ quadruplet with $Q_{B-L} =3$ and both fermion SU(2)$_L$ and SU(2)$_R$ quintuplet do not have operator less than dimension 6.
Note, however, that a DM with $Y \neq 0$ are excluded by DM direct detection due to large 
DM-nucleron scattering cross section via Z-exchange unless some other mechanism suppress 
the cross section.   Therefore we shall focus on the fermion SU(2)$_R$ quintuplet for our study of 
minimal DM since SU(2)$_L$ quintuplet is same as discussed in MDM model in the canonical LR model
~\cite{Heeck:2015qra}.

   \begin{table}[b]
   \caption{Interactions leading to DM decay for scalar(fermion) SU(2)$_{L,R}$ multiplet where ``Reps.'' correspond to representation under SU(2)$_L \times$SU(2)$_R \times$U(1)$_{B-L}$, $Y_{DM}$ is hypercharge of DM, $\tilde H = (i\sigma^2_L) H^T (i \sigma^2_R)$ with $\sigma_{L(R)}^2$ being the second Pauli matrix acting on SU(2)$_{L(R)}$ representation space, and $\tilde H_R = (i \sigma^2_R) H_R^*$. The Lorentz indices are suppressed.}
  \label{tab:DM}
  \begin{minipage}{0.45\hsize}
  \begin{tabular}{llc}  \hline
Reps. \qquad \qquad & Interaction \qquad \qquad & $Y_{DM}$ \\ \hline
\multicolumn{3}{c}{SU(2)$_R$ multiplet scalar $\Phi_R$} \\
(1,2,1) & $\Phi_R H_R^\dagger$ & $0$ \\
(1,3,0) & $\Phi_R \tilde H_R H_R$ & $0$ \\
(1,3,2) & $\Phi_R \tilde H_R \tilde H_R$ & $0$ \\
(1,4,1) & $\Phi_R  \tilde H_R\tilde H_R H_R$ & $0$ \\
(1,4,3) & $\Phi_R  \tilde H_R\tilde H_R \tilde H_R$ & $0$ \\
(1,5,0) & $\Phi_R  \tilde H_R\tilde H_R H_R H_R$ & $0$ \\
(1,5,2) & $\Phi_R  \tilde H_R\tilde H_R \tilde H_R H_R$ & $0$ \\
(1,5,4) & $\Phi_R  \tilde H_R\tilde H_R \tilde H_R \tilde H_R$ & $0$ \\ \hline
\multicolumn{3}{c}{SU(2)$_L$ multiplet scalar $\Phi_L$} \\
(2,1,1) & $\Phi_L  H_R^\dagger \tilde H $ & $1/2$ \\
(3,1,0) & $\Phi_L  H \tilde H $ & $0$ \\
(3,1,2) & $\Phi_L  \bar \ell^c_L \ell_L $ & $1$ \\
(4,1,1) & $\Phi_L  H \tilde H H H_R $ & $1/2$ \\
(4,1,3) & $\Phi_L  \bar \ell^c_L \ell_L H H_R $ & $3/2$ \\
(5,1,0) & $\Phi_L (H \tilde H) (H \tilde H)$ & $0$ \\
(5,1,2) & dim $>$ 5 & $1$ \\
(5,1,4) & dim $>$ 5 & $2$ \\ \hline
\end{tabular}
\end{minipage}
\begin{minipage}{0.45\hsize}
\begin{tabular}{llc}  \hline
Reps. \qquad \qquad & Interaction \qquad \qquad & $Y_{DM}$ \\ \hline
\multicolumn{3}{c}{SU(2)$_R$ multiplet fermion $\Psi_R$} \\
(1,2,1) & $\bar \Psi_R \ell_R^c H^\dagger H$ & $0$ \\
(1,3,0) & $\bar \Psi_R \ell_R H_R $ & $0$ \\
(1,3,2) & $\bar \Psi_R \ell_R^c  H_R $ & $0$ \\
(1,4,1) & $\bar \Psi_R  \ell_R^c H_R \tilde H_R$ & $0$ \\
(1,4,3) & $\bar \Psi_R  \ell_R^c H_R H_R $ & $0$ \\
(1,5,0) & dim $>$ 5 & $0$ \\
(1,5,2) & dim $>$ 5& $0$ \\
(1,5,4) & dim $>$ 5 & $0$ \\ \hline
\multicolumn{3}{c}{SU(2)$_L$ multiplet fermion $\Psi_L$} \\
(2,1,1) & $\bar \Psi_L \ell_L^c H^\dagger H$ & $1/2$ \\
(3,1,0) & $\bar \Psi_L \ell_L H H_R  $ & $0$ \\
(3,1,2) & $\bar \Psi_L \ell_L^c H H_R  $ & $0$ \\
(4,1,1) & $\bar \Psi_L \ell_L^c H \tilde H  $ & $1/2$ \\
(4,1,3) & dim $>$ 5  & $3/2$ \\
(5,1,0) & dim $>$ 5  & $0$ \\
(5,1,2) & dim $>$ 5 & $1$ \\
(5,1,4) & dim $>$ 5 & $2$ \\ \hline
\end{tabular}
\end{minipage}
\end{table}

\section{DM phenomenology}

In this paper we focus on SU(2)$_R$ quintuplets $\Psi^{Q_{B-L}}$ whose electrically neutral 
component provides a DM candidate.   Possible values of $Q_{B-L}$ are 0, 2 and 4, where 
the  corresponding multiplets can be written as
\begin{align}
& \Psi^{0} = (\chi^{++}, \chi^+, \chi^0, \chi^- \chi^{--})^T, \quad \Psi^{2} = (\eta^{+++}, \eta^{++}, \eta^{+}_1, \eta^0, \eta^{-}_2)^T, \nonumber \\
&  \Psi^{4} = (\zeta^{++++}, \zeta^{+++}, \zeta^{++}, \zeta^{+}, \zeta^0)^T.
\end{align}
where subscripts "$+$" etc. denote electric charge of the components and $\chi^0$ is Majorana fermion while the others are Dirac fermion.
We note that $Q_{B-L}=0$ multiplet is discussed in the exact left-right symmetric case in Ref.~\cite{Heeck:2015qra}.

The gauge couplings of a component $\psi^Q$ ($\psi = \chi, \eta, \zeta$) with mass eigenstates of the gauge bosons can be written by
\begin{align}
L \supset & -s_W s_M g_R Q \bar \psi^Q Z^\mu \gamma_\mu \psi^Q + c_M g_R \left( Q - \frac{Q_{B-L}}{2 c_M^2} \right)  \bar \psi^{Q} Z'^\mu \gamma_\mu \psi^{Q} \nonumber \\
&+ c_W s_M g_R Q \bar \psi^{Q} A^\mu \gamma_\mu \psi^{Q} + \frac{g_R}{\sqrt{2}} ( c_{2 m} \bar \psi^{Q+1} W'^{+\mu} \gamma_\mu \psi^{Q} + h.c. ),
\label{eq:int}
\end{align}
where $c_{2m} = \sqrt{(2+m+1)(2-m)}$ with $m = Q -Q_{B-L}/2$.

\subsection{Mass splitting}

\begin{figure}[t] 
\begin{center}
\includegraphics[width=60mm]{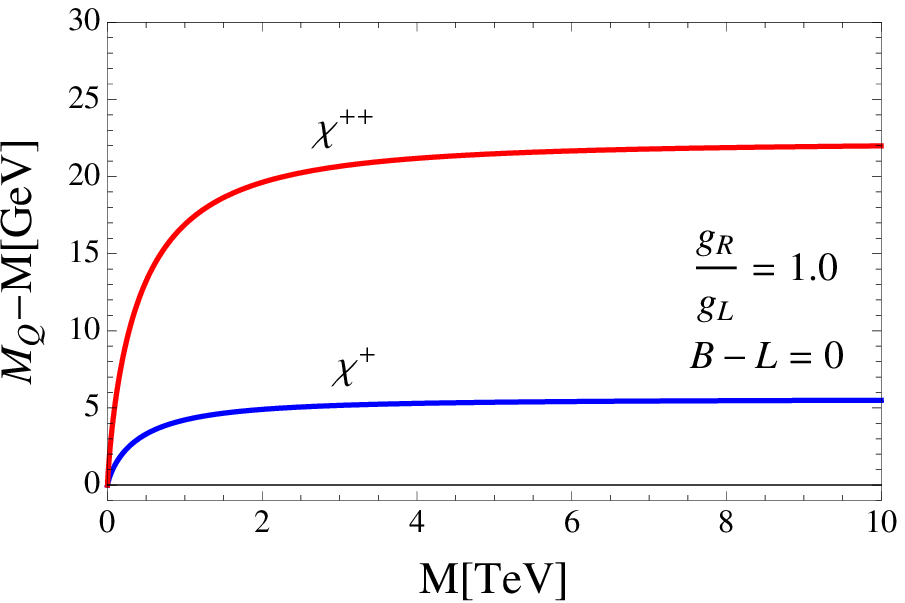} 
\includegraphics[width=60mm]{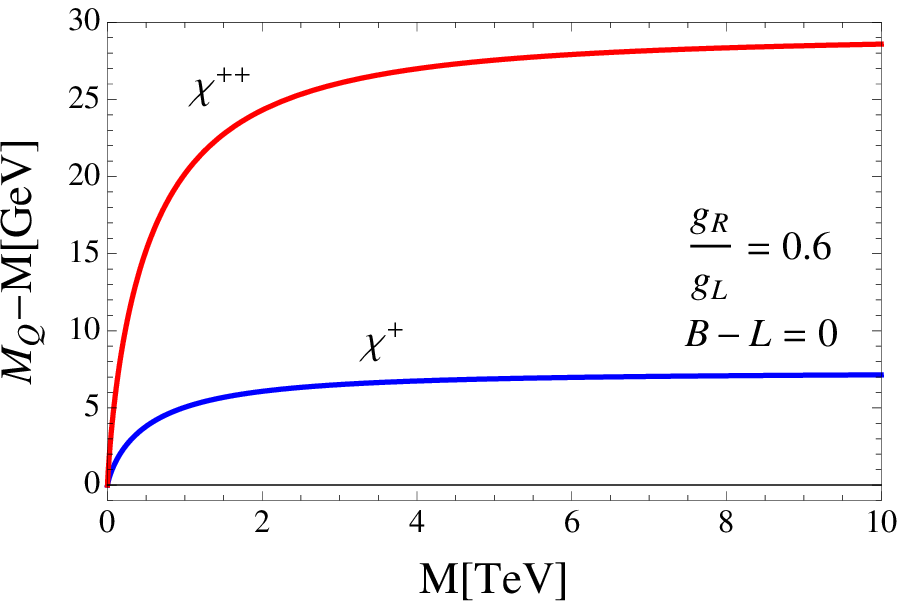}
\includegraphics[width=60mm]{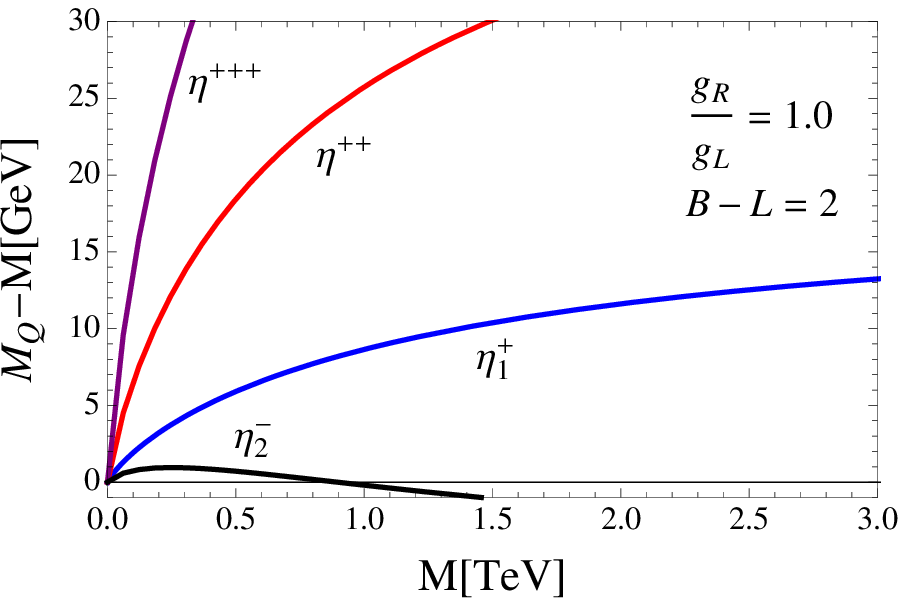}
\includegraphics[width=60mm]{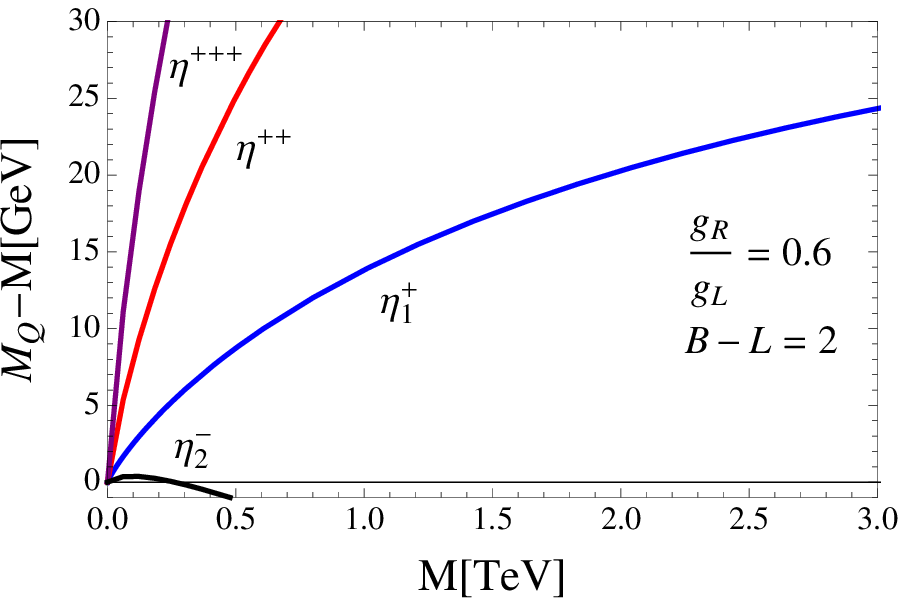}
\includegraphics[width=60mm]{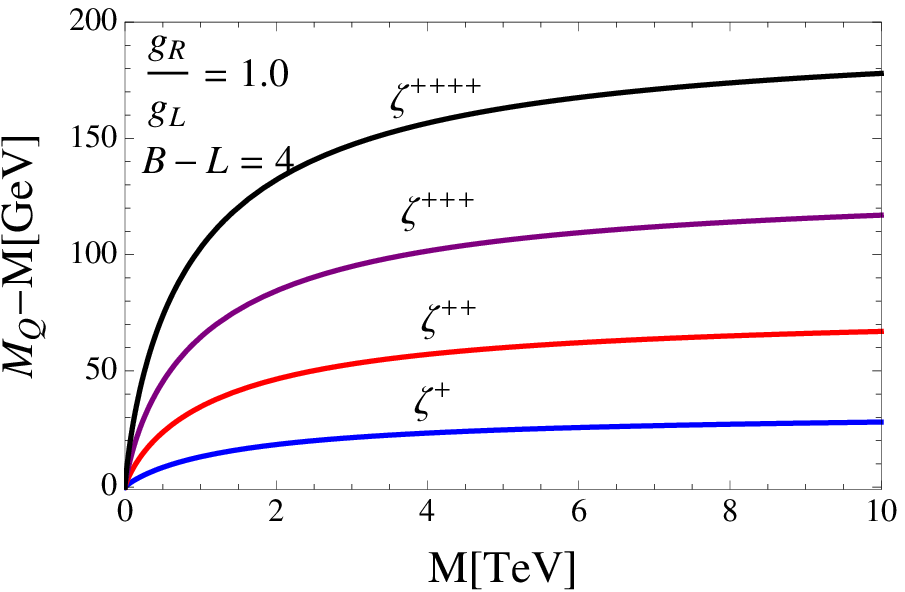} 
\includegraphics[width=60mm]{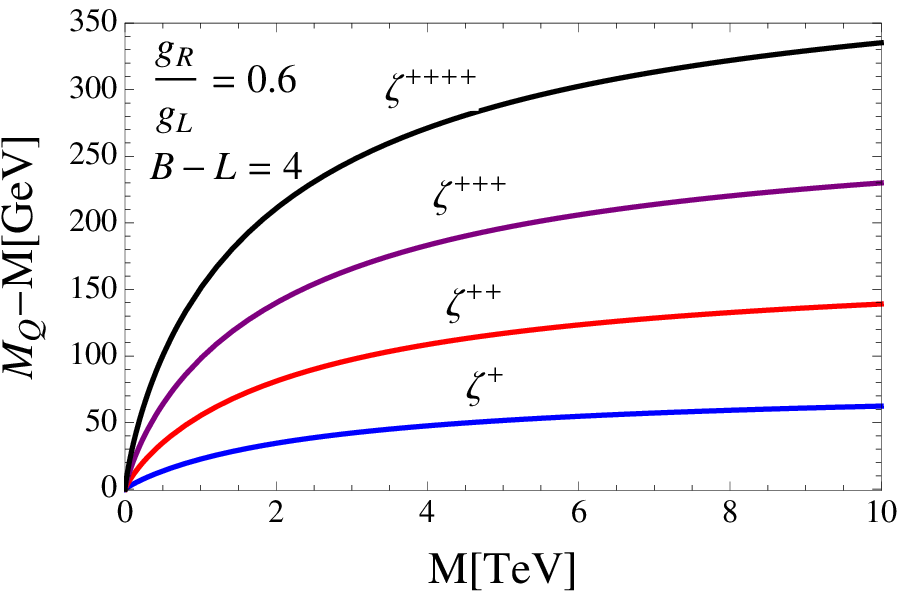} 
\caption{ Mass difference between neutral and charged components of right-handed quintuplet where $B-L= \{ 0, 2, 4 \}$ and $g_R/g_L = \{ 1, 0.6 \}$.}
\label{fig:DeltaM1}
\end{center}
\end{figure}

The mass splitting between charged components and the neutral component in a given multiplet 
can be obtained by calculating radiative correction where the gauge bosons propagate inside loop 
diagrams.   We then find the formula of the mass splitting as 
\begin{align}
M_Q -M_0 \simeq & \frac{g_R^2}{(4 \pi)^2} M [Q (Q - Q_{B-L}) f(r_{W'}) - c_M^2 Q \{ Q - Q_{B-L}/ c_M^2 \} f(r_{Z'})  \nonumber \\
& \qquad  \qquad - s_W^2 s_M^2  Q^2 f(r_{Z}) - c_W^2 s_M^2  Q^2 f(r_\gamma) ]
\end{align} 
where $Q$ is electric charge, $r_X = m_X/M$ and $f(r) \equiv 2 \int_0^1 dx (1+x) \log [x^2 + (1-x)r^2]$.
We note that $Q_{B-L}=0$ provides same formula as in Ref.~\cite{Heeck:2015qra}.
Fig.~\ref{fig:DeltaM1} shows the mass difference $M_Q - M$ for $m_{W'} = 2$ TeV and 
$g_R/g_L = 1.0 (0.6)$ where $M_Q$ and $M$ are masses of component with charge $Q$ and of 
DM respectively.
We find that the mass splitting $M_Q-M$ is always positive for $Q_{B-L}=0$ which is qualitatively different from Ref.~\cite{Heeck:2015qra} where the $M_Q-M$ becomes negative when DM mass 
$M$ is larger than $\sim$1.8(4.5) TeV for $m_{W'} = 2 (5)$ TeV.
This difference comes from mass relation between $m_{Z'}$ and $m_{W'}$; $m_{Z'}/m_{W'}$ 
is smaller in our case when the same $g_R/g_L$ value is applied.
For the multiplet with $Q_{B-L}=2$, the second singly charged component $\eta_2^-$ becomes 
lighter than the neutral component for $M \gtrsim 1 (0.6) $ TeV.
The $Q_{B-L}=4$ multiplet also provides positive $M_Q - M$ for all $M$ value where the mass 
splitting is larger than the case of $Q_{B-L}=0$.

\subsection{Relic density}

\begin{figure}[t] 
\begin{center}
\includegraphics[width=60mm]{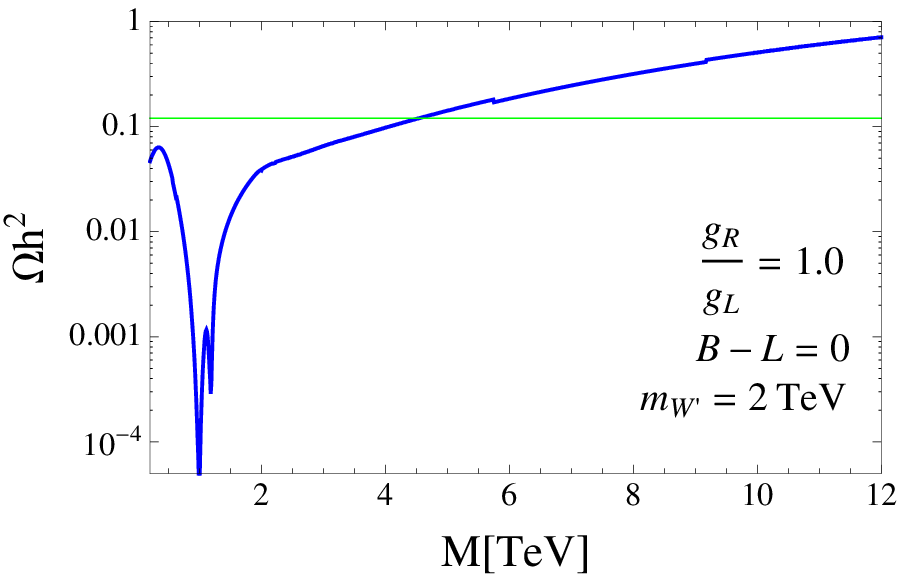}
\includegraphics[width=60mm]{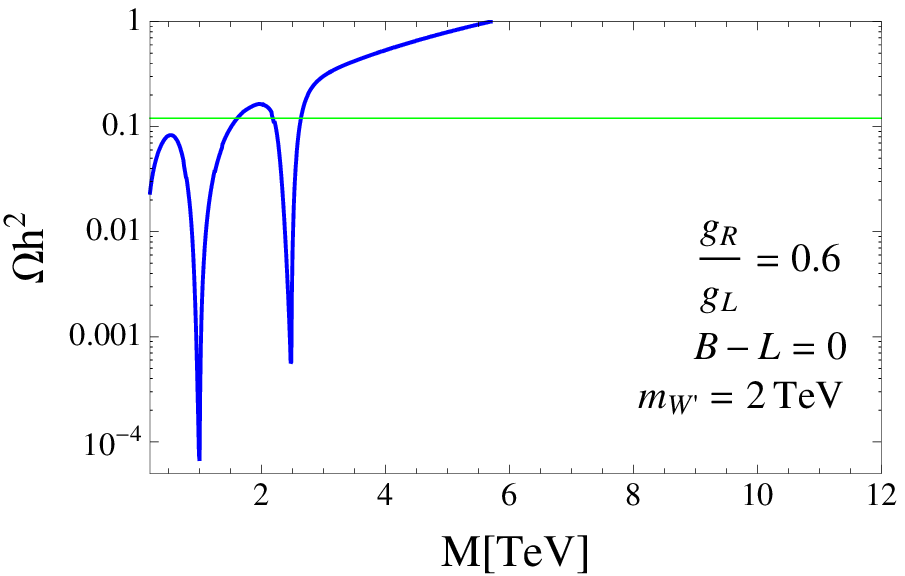}
\includegraphics[width=60mm]{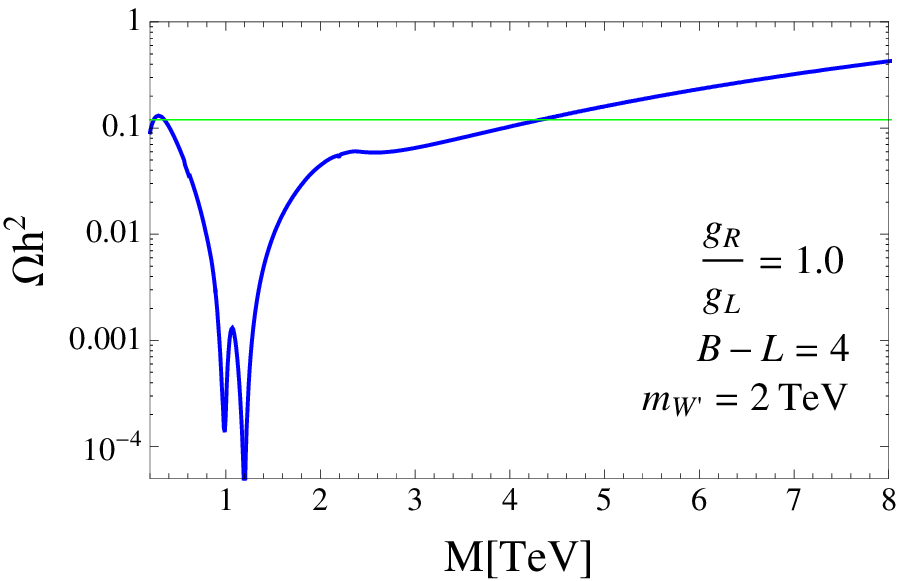}
\includegraphics[width=60mm]{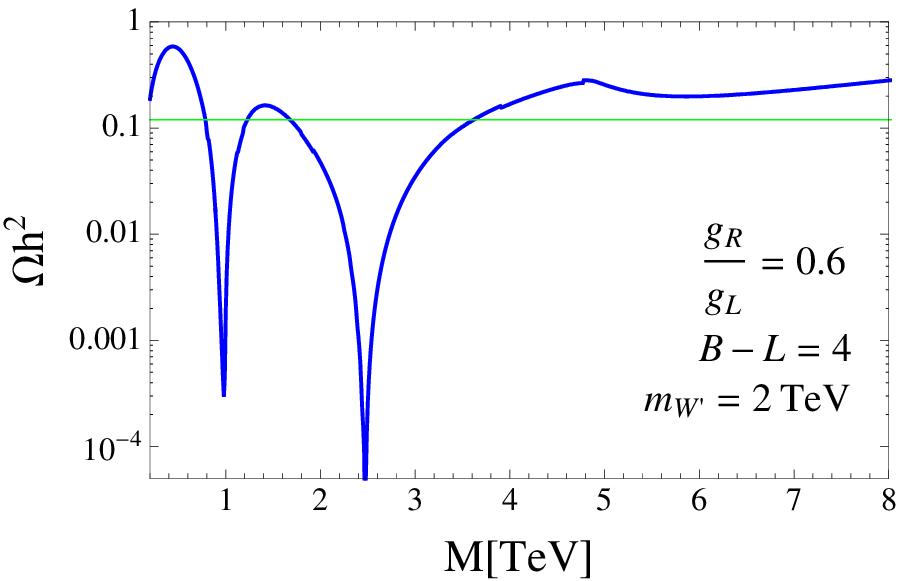}
\caption{ Relic density $\Omega h^2$ for fermionic SU(2)$_R$ quintuplet DM with $B-L=0$ and $4$ where the left(right) panel shows $g_R/g_L = 1.0(0.6)$. }
\label{fig:RD}
\end{center}
\end{figure}

Thermal relic density of DM is numerically calculated using { \tt micrOMEGAs 4.1.5 }  \cite{Belanger:2014vza} to solve the Boltzmann equation by implementing relevant interactions 
providing (co)annihilation processes of DM.
Here the (co)annihilation processes of DM are induced by gauge interactions in Eq.~(\ref{eq:int}).
We show the estimated relic density by blue lines in Fig.~\ref{fig:RD} for $Q_{B-L}=0$ and 4 with $m_{W'} =2$ TeV where the left(right) panels correspond to $g_R/g_L = 1.0(0.6)$.
It is compared to the value measured by Planck~\cite{Ade:2013kta}, $\Omega h^2 = 0.1199 \pm 0.0027$, indicated by green horizontal line in the plots.
Note that the case of $Q_{B-L}=2$ is not shown in Fig.~\ref{fig:RD} since it can not provide observed relic density in the region of $M$ where the neutral component is the lightest.
The plots show the resonance effects around $M \sim m_{W'}/2$ and $M \sim m_{Z'}/2$ where (co)annihilation cross sections become large decreasing relic density.
We find that the relic density tends to larger for smaller $g_R/g_L$ since the (co)annihilation cross section is suppressed by following effects: 
({\it i}) the heavier $Z'$ for the smaller 
$g_R/g_L$,  and ({\it ii}) the smaller coupling constant $g_R$ compared with $g_L$.
Also $Q_{B-L}=4$ multiplet provides larger relic density than $Q_{B-L}=0$ multiplet since (a) DM is Dirac fermion for $Q_{B-L}=4$ (b) larger mass splitting suppress the coannihilation effect. 
The masses of DM giving observed relic density are summarized in third column of Table.~\ref{tab:pheno} for each case.
We note that SU(2)$_R$ quintuplet can provide observed relic density with $m_{W'} = 2$ TeV unlike the case of the exact left-right symmetric model in Ref.~\cite{Heeck:2015qra} due to different mass splitting relations. Furthermore, when we apply $g_R/g_L =0.6$ the required mass of DM can be as light as ${\cal O}(1)$ TeV which is light compared to SU(2)$_L$ quintuplet in MDM model~\cite{Cirelli:2007xd}. 

\subsection{DM nucleon scattering cross section}

For the multiplet with $Q_{B-L} \neq 0$, DM can interact with nucleon by exchanging $Z'$ boson. 
The DM-$Z'$ coupling is obtained from Eq.~(\ref{eq:int}) while $Z'$ couples to right-handed $u$- and $d$-type quark currents with couplings $c_M g_R (2/3-1/(6 c_M^2))$ and $c_M g_R (-1/3-1/(6 c_M^2) )$ respectively.
We then estimate DM-nucleon scattering cross section for $\zeta^0$ to investigate constraint from DM direct detection experiment.
The spin independent elastic scattering cross section of DM and nucleon $N$ can be calculated as 
\begin{equation}
\sigma_{\zeta^0 N} \simeq \frac{4 g_R^2 g_{NNZ'}^2}{ \pi c_M^2} \frac{1}{m_{Z'}^4} \frac{m_N^2 M^2}{(m_N+ M)^2}
\end{equation}
where $m_N$ is nucleon mass, $g_{ppZ'} = c_M g_R (1/2-1/(4 c_M^2) )$ and $g_{nnZ'} = - g_R/(4 c_M) $. 
Since our DM is much heavier than nucleon the cross section is almost independent of DM mass.
We then find that the values of cross section averaged by nucleon $(\sigma_{\zeta^0 p} + \sigma_{\zeta^0 n})/2$ are 
$6.8 \times 10^{-44}$cm$^2$ and $1.1 \times 10^{-44}$cm$^2$ for $g_R/g_L=$ 1.0 and 0.6 respectively.
These cross sections are compared to current constraint given by LUX experiment~\cite{Akerib:2013tjd} where the upper limits are given in seventh column of Table.~\ref{tab:pheno} for each DM mass. 
Thus the cases of $g_R/g_L = 1.0$ and of $g_R/g_L=0.6$ with  $M \lesssim 1 $TeV are excluded unless some cancellation mechanism work while that of $g_R/g_L = 0.6$ with $M \gtrsim 1$ TeV are allowed.
Interestingly, the scattering cross sections for proton and neutron are significantly different which are summarized in sixth column of Table.~\ref{tab:pheno}. 
For the case of $Q_{B-L}=0$, DM interacts with nucleon at one-loop level exchanging $W'$ boson and the scattering cross section is much smaller than the current constraint.

   \begin{table}[b]
   \caption{DM mass giving relic density $\Omega h^2 =0.1199$, mass difference $M_{Q} - M$, cross section for $pp \rightarrow \psi^Q \psi^{Q'}$ at 13(14) TeV where $\psi^Q \psi^{Q'}$ includes all possible combination including charged component,  $\sigma_{\rm DM-N}$ denotes a DM-nucleon scattering cross section where the value outside(inside) bracket is for proton(neutron), and $\sigma_{\rm Lux}$ is current limit by Lux~\cite{Akerib:2013tjd}.}
  \label{tab:pheno}
  \begin{tabular}{ccccccc}  \hline
  $B-L$ & \, $g_R/g_R$ \, & \, $m_{DM}$[TeV] \, & \, $\Delta M$[GeV] \, & \quad $\sigma_{\psi^Q \psi^{Q'}} $[fb] \quad & \quad  $ \sigma_{\rm DM-N}$[cm$^2$] \quad & \quad $\sigma_{\rm Lux}$[cm$^2$] \quad \\ \hline
  0 & 1 & 4.54 & 5.34 & $\ll 10^{-2}$ & $\ll 10^{-45}$ & \quad $\sim 57. \times 10^{-45}$   \\ \hline
  0 & 0.6 & 1.61 & 5.79 & 0.11(0.18) & $\ll 10^{-45}$ & \quad $\sim 18. \times 10^{-45}$ \\
     &       & 2.18 & 6.18 &  0.034(0.069) & $\ll 10^{-45}$ & \quad  $\sim 26. \times 10^{-45}$  \\
     &       & 2.64 & 6.39 & $\ll 10^{-2}$ &  $\ll 10^{-45}$ & \quad $\sim 30. \times 10^{-45}$  \\ \hline
  4 & 1   &  0.244 & 5.07 & 2010(2380) &  $1.9(12.) \times 10^{-44}$ & \quad  $\sim 3.1 \times 10^{-45}$   \\
     &      &  0.356 & 6.71 & 1420(1740)  &  $1.9(12.) \times 10^{-44}$  & \quad $\sim 4.3 \times 10^{-45}$ \\
     &      &   4.32 & 23.7 & $\ll 10^{-2}$ &   $1.9(12.) \times 10^{-44}$ & \quad $\sim 52. \times 10^{-45}$   \\ \hline
  4 & 0.6  &  0.785 & 19.1 & 371.(460) &  $6.8(15.) \times 10^{-45}$ & \quad $\sim 8.9 \times 10^{-45}$ \\ 
     &      &  1.23 & 25.9 & 1.44(2.05) & $6.8(15.) \times 10^{-45}$ & \quad $\sim 14. \times 10^{-45}$ \\
     &      &  1.66 & 31.1 & 0.173(0.298) & $6.8(15.) \times 10^{-45}$ & \quad $\sim 19. \times 10^{-45}$ \\
     &      &  3.62 & 45.7 & $\ll 10^{-2}$ & $6.8(15.) \times 10^{-45}$ & \quad $\sim 45. \times 10^{-45}$ \\ \hline
\end{tabular}
\end{table}

\subsection{Implication to collider physics}

The components in SU(2)$_R$ quintuplet can be produced at the LHC through gauge interactions $pp \rightarrow V \rightarrow \bar \psi^{Q} \psi^{Q'}$ where $V$ can be $\gamma$, $Z$, $W'$ or $Z'$ according to interactions in Eq.~(\ref{eq:int}). 
Since DM just appears as missing transverse energy $E_T$ we consider production processes which include charged components;  $p p \rightarrow \bar \psi^Q  \psi^Q$, $p p \rightarrow \bar \psi^Q \psi^{Q\pm1}$, and $p p \rightarrow \bar \psi^\pm \psi^{0}$ where $\psi = \chi, \eta$ and $Q \neq 0$.
Here we estimate the production cross section numerically by CalcHEP~\cite{Belyaev:2012qa} utilizing the code with {\tt CTEQ6L} PDF~\cite{Nadolsky:2008zw} for $\sqrt{s}=13$ and $14$ TeV.
We show the production cross sections in 5th column of Table.~\ref{tab:pheno} where all production modes are summed over.
The cross section can be larger than $0.1$ fb for $M \lesssim 2$ TeV which would be within reach of the LHC.

The charged components can decay as $\psi^Q \rightarrow \psi^{Q \mp 1} M^\pm (M^\pm = \pi^\pm, K^\pm, {\rm etc.})$ and $\chi^Q \rightarrow \chi^{Q \mp 1} \bar q' q$ where off-shell $W'$ converts into $M^\pm$ and light quarks $\bar q' q$ assuming heavy right-handed neutrinos.
Since the mass splittings are $\Delta M >$ 5 GeV in our case, $\psi^Q \rightarrow \psi^{Q \mp 1} \bar q' q$ provide dominant contribution of decay width which is given by 
\begin{align}
\Gamma(\psi^Q \rightarrow \psi^{Q - 1} \bar q' q) \simeq  N_c c_{2m}^2 \frac{g_R^4}{120 \pi^3}  \frac{ \Delta M^5 }{m_{W'}^4}
\end{align}
where $Q$ is taken to be positive here. 
We find that the lifetime of the charged components are less than $1 {\rm cm}/c$ and they would decay within the non-detector region.
Thus the signal of quintuplet production is "jets + missing $E_T$" in our case.

Here we briefly discuss the diboson excess in our model. Calculating with CalcHEP,
the production cross section for $pp \to W'$ is given as $\sigma (pp \to W') \simeq 220 (g_R/g_L)^2$ fb for $m_{W'} = 2$ TeV, which is similar to other SU(2)$_L \times $SU(2)$_R \times$U(1)$_{B-L}$ models~\cite{Awasthi:2015ota, Heeck:2015qra,Hisano:2015gna, Cheung:2015nha, Dobrescu:2015qna, Aguilar-Saavedra:2015rna, Gao:2015irw, Brehmer:2015cia, Cao:2015lia, Dev:2015pga, Deppisch:2015cua, Goncalves:2015yua}. 
On the other hand the branching fraction for $W' \to W Z$ would be different since we have additional SU(2)$_R$ quintuplet fermion.
The $W'$ can decay into quintuplet states where the partial width is given by 
\begin{equation}
\Gamma \left( W' \to \sum_{i, j} \bar \psi_i \psi_j  \right) = \left(\sum_{ij} C_{ij}^2 \right) \frac{g_R^2}{12 \pi} m_{W'} \left( 1 + \frac{2 M^2}{m_{W'}^2} \right) \sqrt{1- \frac{4 M^2}{m_{W'}^2}} 
\end{equation}
where $\psi_i$ represents a component in quintuplet, and 
\[
\sum_{ij} C_{ij}^2 = \left\{ 
       \begin{array}{cc} 
    (\sqrt{2})^2 + \frac{1}{2} (\sqrt{3})^2 &=  \frac{7}{2} \quad {\rm for} \, \, B-L=0 ~~{\rm quintuplet} , \\
     (\sqrt{2})^2 + (\sqrt{3})^2 +(\sqrt{3})^2 + (\sqrt{2})^2  &= 10 \quad {\rm for} \, \, B-L = 4 ~~{\rm quintuplet}, 
       \end{array} \right.
\]
where $1/2$ factor appears in second term of LHS for B-L=0 since the neutral component is Majorana fermion.
We assumed all the components $\psi_i$ have the common mass $M$, ignoring mass difference in the quintuplet for simplicity.  The other partial decay widths are given by
\begin{align}
\Gamma \left( W' \to \sum \bar q q' \right)  &= \frac{3 g_R^2}{16 \pi} m_{W'} \nonumber \\
\Gamma (W' \to W Z) & = \Gamma (W' \to W h) = \frac{g_L^2}{192 \pi} \sin^2 \xi \frac{m_{W'}^5}{m_{W}^4}
\end{align}
where $\sum \bar q q'$ indicates sum of possible combinations of SM quarks, $\xi$ is 
the $W$-$W'$ mixing angle in Eq.(\ref{eq:mixing1}) and all the quark masses are omitted.
We find that branching fraction of $W' \to \sum_{i, j} \bar \psi_i \psi_j  $ is dominant when 
$2 M < m_{W'}$.  In this case  a value of
$\sin \xi$ should be larger than the value in a left-right model without extra multiplet in order to obtain
$\sigma (pp\to W') BR(W' \to WZ) \sim 10$ fb for explaining the diboson excess; 
 $\sin \xi \sim 3(2) \times 10^{-3}$ and $\sim 5(3) \times 10^{-3}$  are required for the cases of $B-L =0$ and 4 with $2 M < m_{W'}$ and $g_R/g_R = 1.0(0.6)$ while 
$\sin \xi \sim 2(1) \times 10^{-3}$ is required for $2 M > m_{W'}$ and $g_R/g_R = 1.0(0.6)$ as the other left-right models.

\section{Summary and discussion}

We have studied a DM model with SU(2)$_L \times$SU(2)$_R \times$U(1)$_{B-L}$ gauge symmetry where the 2 TeV $W'$ boson from the extended gauge symmetry is a potential explanation of di-boson excess in ATLAS and $eejj$ excess in CMS.
The gauge symmetry is broken by VEVs of SU(2)$_R$ doublet and SU(2)$_{L,R}$ bi-doublet scalar field where the scalar sector is simplified loosening the exact left-right exchange symmetry.
Then we have classified a new fermion and scalar SU(2)$_{L,R}$ multiplet which can be stable in cosmological timescale based on the idea of minimal dark matter.

As a new candidate of DM, we focused on three fermion SU(2)$_R$ quintuplets with $B-L = 0, 2$ and 4.
We investigated mass splitting between neutral and charged components of the quintuplets, relic density of DM, and DM-nucleon scattering for each multiplets adopting $m_{W'} = 2$ TeV.
For the mass splitting of $B-L=0$ quintuplet, we find that neutral component is always lightest where it is qualitatively different from the exact left-right symmetric case.
We have also shown that the neutral component is always the lightest for $B-L=4$ while a charged component in $B-L=2$ quintuplet becomes the lightest when DM mass is $M \gtrsim 1.0(0.6)$ TeV for $g_R/g_L = 1.0(0.6)$.
The observed relic density can be obtained for the $B-L=0$ and 4 quintuplets but the $B-L=2$ quintuplet give smaller relic density in the region where neutral component is the lightest.
Furthermore the investigation of DM-nucleon scattering for $B-L =4$ excludes cases of $g_R/g_L=1.0$ and $g_R/g_L=0.6$ with $M  \lesssim 1$ TeV since the scattering cross section via $Z'$ exchange is larger than current limit. 
Thus the values of DM mass are predicted to be $4.5(1.6, 2.2, 2.6)$ TeV for $B-L=0$ with $g_R/g_L=1.0(0.6)$ and $\{1.2, 1.7, 3.6 \}$ TeV for $B-L=4$ with $g_R/g_L=0.6$ respectively.
We then show that charged components can be produced at the LHC 13 and 14 TeV with total cross section around $0.1$ to $2.1$ fb when DM mass is relatively light as $1.2 $ to $1.7$ TeV.
The charged components predominantly decay into an other charged (neutral) component with 1-unit charge difference and light quarks, which lead the signal of the quintuplet production as jets plus missing $E_T$.
Further analysis of the signal is left as future work.

\vspace{1cm}

\noindent{\bf Acknowledgments}
We are grateful to Ayon Patra for discussions on the left-right sym- metric models. 
This work is supported in part by National Research Foundation of Korea (NRF) Research Grant NRF-2015R1A2A1A05001869, and by SRC program of NRF Grant No. 20120001176 funded 
by MEST through Korea Neutrino Research Center at Seoul Na- tional University (PK).

\end{document}